\documentclass[aps,twocolumn,superscriptaddress,amsmath,showpacs]{revtex4}
\usepackage{graphicx}
\usepackage{bm}
\usepackage{amsmath}
\begin{document}

\title{Exploring Quantum Phase Transitions with a Novel Sublattice Entanglement Scenario}

\author{Yan Chen}
\affiliation{Department of Physics \& Center of Theoretical and
Computational Physics, The University of Hong Kong, Pokfulam Road,
Hong Kong, China}
\author{Z. D. Wang}
\affiliation{Department of Physics \& Center of Theoretical and
Computational Physics,  The University of Hong Kong, Pokfulam
Road, Hong Kong, China}
\author{F. C. Zhang}
\affiliation{Department of Physics \& Center of Theoretical and
Computational Physics,  The University of Hong Kong, Pokfulam
Road, Hong Kong, China}

\begin{abstract}
We introduce a new measure called reduced entropy of sublattice to
quantify entanglement in spin, electron and boson systems. By
analyzing this quantity, we reveal an intriguing connection between
quantum entanglement and quantum phase transitions in various
strongly correlated systems: the local extremes of reduced entropy
and its first derivative as functions of the coupling constant
coincide respectively with the first and second order transition
points.  Exact numerical studies merely for small lattices reproduce
several well-known results, demonstrating that our scenario is quite
promising for exploring quantum phase transitions.

\end{abstract}


\maketitle

Quantum entanglement, a key concept in quantum information theory
that has no classical counterpart~\cite{Wootters98},
 has been recognized to play also an important role in the study
of quantum many-particle physics
~\cite{Aeppli}. However, how to quantify appropriately
entanglement in many-body systems has been a challenging question
for a long time. So far, a number of interesting theoretical
attempts have been made on various notions of entanglement in
several one-dimensional (1D) correlated systems as well as their
possible links to quantum phase transitions (QPTs) that occur at
absolute zero temperature with the change of coupling
parameters~\cite{Sachdev,Arnesen01,Connor01,nature2002,Wang02,GVidal03,Cirac04}.
Most of these studies focused mainly on the spin chains, where the
spin-spin concurrence~\cite{Wootters98} was  adopted to describe
(two-particle) entanglement, which can be calculated from
correlation functions~\cite{Wang02}. Since the correlation
function decays rapidly, the concurrence is nonzero only between
two closer sites and thus the information on QPTs abstracted from
the concurrence is rather limited~\cite{Chen04}. Scaling behavior
of entanglement between a block of contiguous spins and the rest
of the system in a spin chain was also studied around the quantum
critical point~\cite{GVidal03}, but the idea of the
contiguous-spins entanglement is merely applicable to few 1D
models without classifying the transition order. Therefore, it is
of significance to attempt other more applicable measures of
entanglement for correlated systems so that the nonlocal nature of
QPTs can be better captured.

In this paper, by introducing a well-defined reduced entropy of
sublattice as a measure of entanglement, which is an extensive
quantity, we have found a distinct connection between quantum
entanglement and QPTs in such an intriguing way: the local
extremes of reduced entropy and its first derivative as functions
of the coupling constant coincide respectively with the first and
second order transition points of QPTs. This finding is remarkable
as (i) it implies that the
 reduced entropy of sublattice plays a crucial role in QPTs, similar
to that of the thermal entropy in classical phase transitions; and
(ii) the analysis of reduced entropy enables us to identify and
classify unambiguously QPT points for a wide class of systems.

\begin{figure}[b]
\includegraphics[width=8.8cm]{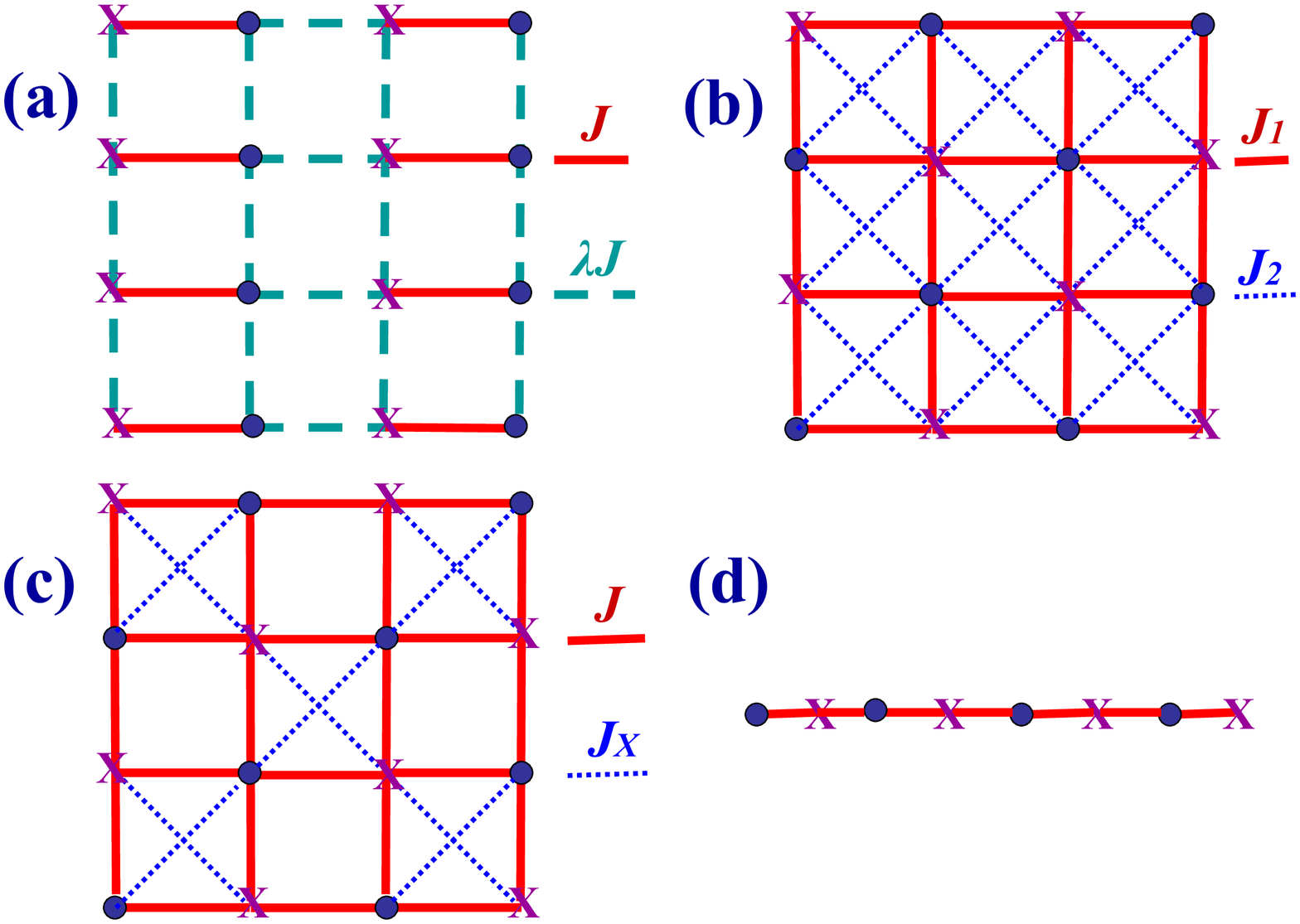}
\caption{\label{Figure 1} Schematic configurations of the four
studied strongly correlated systems. a) Coupled dimer
antiferromagnet, b) 2D frustrated $J_1$-$J_2$ system, c)
checkerboard antiferromagnet and d) 1D extended Hubbard model. Two
subsystems are denoted as $X$ (crosses) and $\bullet$ (solid
symbols) in the figures.}
\end{figure}

While we have studied more than ten strongly correlated systems with
the same conclusion, here we first analyze the best understood
transverse-field quantum Ising chain as a benchmark, and then
present our main results for three types of two-dimensional (2D)
spin-1/2 systems and one type of electron system: a coupled dimer
antiferromagnet (Fig.1a), a frustrated $J_1$-$J_2$ model (Fig.1b), a
checkerboard antiferromagnet (Fig.1c), and a 1D half-filled extended
Hubbard model (Fig.1d). The corresponding Hamiltonians read
\begin{eqnarray}
&H_{\alpha}&= \sum_{ij}J_{ij}^{\alpha} \mbox{\boldmath$S$}_{i}
\mbox{\boldmath$S$}_{j},
\\
&H&=-\sum_{\langle ij \rangle \sigma }c_{i\sigma }^{\dagger
}c_{j\sigma}+U\sum_{i}n_{i\uparrow }n_{i\downarrow } +
V\sum_{\langle ij \rangle}n_{i}n_{j},
\end{eqnarray}
where $\mbox{\boldmath$S$}_{i}$ is the spin operator and $\alpha$ is
a  model label. The coupling parameters in all models are explicitly
shown in Fig.1. For the coupled dimer antiferromagnet model
$(\alpha=a)$, the anisotropy parameter is denoted by $\lambda$.
$J_1$ and $J_2$ correspond to the nearest-neighbor and
next-nearest-neighbor exchange couplings in the 2D frustrated
Heisenberg Hamiltonian $(\alpha=b)$. In the checkerboard
antiferromagnet spin system $(\alpha=c)$, the Hamiltonian contains
nearest-neighbor couplings along the nearest-neighbor coupling $J$
and the diagonal link $J_{\times}$. As for the 1D half-filled
extended Hubbard model, $c^\dagger_{j \sigma}$ and $c_{j \sigma}$
are creation and annihilation operators of electron with spin
$\sigma$ at site $j$, respectively, $U$ and $V$ represent the
on-site repulsion/attraction and nearest-neighboring interaction.

In an earlier paper, Vidal {\em et al}.~\cite{GVidal03} studied
the ground state entanglement between a block of  contiguous spins
 and the rest of the system. Their approach is, however, not
applicable to the above four systems and many others. In the
present work,  it is in fact crucial to choose one size-$N$
sublattice $R_N$ {\it reduced} from the whole lattice in such a
preferable way (i) the number of connecting bonds between the
chosen sublattice and the rest ($B_N$) of the system  is maximized
so that the correlation effect between the two parts can be best
revealed; and (ii) the structure of sublattice $R_N$ should be the
same as (or, if impossible, most similar to) that of the original
lattice, in the same spirit of the real-space renormalization
idea. Under such choice, a  new appropriate measure of
entanglement for correlated systems, i.e. the  reduced entropy of
sublattice, is naturally given by $S_N = - \rm{tr} \left( \rho_{N}
\log_2 \rho_{N} \right)$, where $\rho_N$ is the reduced density
matrix by the partial trace of the density matrix over $B_N$
~\cite{Wootters98}. Being different from some conventional
quantities (e.g., the spin susceptibility) which are only related
to a two-body correlation function, this reduced entropy of
sublattice contains all the contributions including two-body,
three-body, and even $N/2$-body correlation functions. Since this
reduced entropy can capture essentially the nonlocal nature of
many-particle correlation, particularly around the transition
points of QPTs, we naturally anticipate that it should have a
close connection to QPTs, as we will elaborate later. In the
numerical calculations (except for the first example addressed
below), we employed Lanczos algorithms with periodic boundary
conditions to calculate exactly the ground state $|\Psi_g\rangle$,
from which we obtained $\rho_N \equiv \rm{tr}_{B_N}
|\Psi_g\rangle\langle\Psi_g|$ and then $S_N$, with the lattice
size $N$ and choice of the sublattice $R_N$ (the lattice of
$X$-points (crosses)) used in our calculation being illustrated in
Fig.1.

As a benchmark illustration, we first analyze an exactly solvable
model intensively studied
before~\cite{Sachdev,nature2002,GVidal03}, i.e., the
transverse-field quantum Ising chain with the Hamiltonian as $H=
-\sum_{i=1}^{L} \left(S_i^x S_{i+1}^x + \lambda S_i^z \right)$. It
is well known that this system undergoes a second order phase
transition between an ordered ferromagnetic state and a quantum
paramagnetic state at $\lambda_c =1$. As depicted in Fig. 2 (with
the even sites sublattice), $S_N$ monotonically decreases as
$\lambda$ changes from 0 to 2, while the first derivative
$d(S_{N}/N)/d\lambda$ shows explicitly a local minimum at
$\lambda=\lambda_c$, indicating clearly that this local extreme
corresponds to a critical point of the second-order QPT. This
one-to-one correspondence between the reduced entropy and QPTs has
also been tested and justified in the study of the Anderson
localization model describing disorder-driven metal-insulator
transition~\cite{footnote1}. In fact, from our numerical results
including those to be addressed later/elsewhere, we state more
firmly that the extreme of the first derivative of reduced entropy
in a finite system corresponds
to the transition point of second-order
QPTs. Moreover, as shown clearly in Fig. 2, as the lattice size
increases from $N=10$ to $1280$, the determined critical point is
unchanged, and the curves of the renormalized first derivative
(and reduced entropy) converge very rapidly (for $N\geq 160$, all
curves merge to one within the present
resolution)~\cite{footnote2}. Therefore, our
 reduced entropy of sublattice indeed captures essentially the nonlocal
nature of many-particle correlation and reveals QPTs with
negligible (or very weak) finite size effect, which is a
remarkable advantage in exploring QPTs in other complicated
systems.
\begin{figure}[t]
\includegraphics[width=6.2cm]{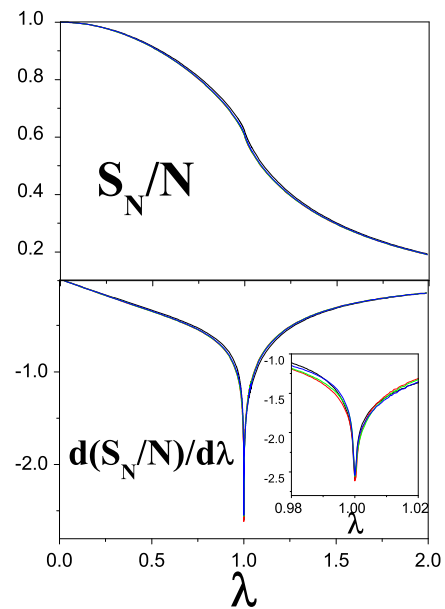}
\caption{\label{Figure 2} The ground state renormalized reduced
entropy $S_N/N$ (a)  and its first derivative versus the external
parameter $\lambda$ (b). The curves correspond to different
lattice sizes $N=10, 40, 160, 1280$. }
\end{figure}

We now use the above scenario to look into a coupled Heisenberg
dimer Hamiltonian with two spins per unit cell, a typical model
for illustration of novel QPTs in the literature~\cite{Senthil04}.
The transition may be tuned by varying a dimensionless parameter
$\lambda$ ($0<\lambda<1$). As shown in Fig.1 (a), the bonds
represented by the solid-line  form the coupled dimers while the
dashed-line represents the coupling between the neighboring
dimers. In the case of $\lambda \ll 1$, the ground state is a
paramagnetic state consisting of a product of decoupled dimers, in
each of which the spins pair into a valence bond singlet. At
$\lambda$=$1$, the system corresponds to the square lattice
antiferromagnet; its ground state has a long-range magnetic Neel
order. Obviously, these two limits cannot be connected
continuously and a transition point is expected between them.
Recently, a field theory study indicated that the transition
between the two phases is of second-order~\cite{Senthil04}, and
the critical point was estimated around $ 0.5$ from a Monte Carlo
simulation~\cite{Matsumoto02}. For the sublattice specified in
Fig.1(a), the reduced entropy $S_N$ and its first derivative as
functions of coupling $\lambda$ are shown in Fig. 3(a). $S_N$
monotonically decreases as $\lambda$ changes from 0 to 1, while
the first derivative $d(S_{N}/N)/d\lambda$ shows a clear local
minimum around $\lambda \sim 0.4$, which is now identified to be a
critical point of the second-order QPT, in good agreement with the
well-accepted result.
\begin{figure}[t]
\includegraphics[width=6.8cm]{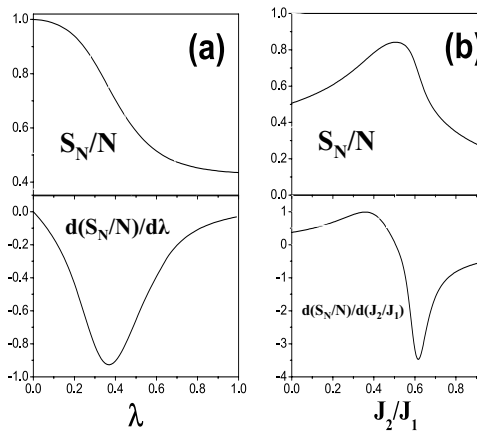}
\caption{\label{Figure 3} The ground state renormalized reduced
entropy $S_N/N$ and its first derivative versus the external
parameter. a) Coupled Dimer Antiferromagnet, b) 2D frustrated
$J_1$-$J_2$ system.}
\end{figure}

For the 2D frustrated $J_1$-$J_2$ Heisenberg model (Fig.1(b)), it
is known that the ground state of the model possesses the
conventional Neel long range order at $J_2/J_1 \ll 1$ and has a
collinear antiferromagnetic quasi-long range order if $J_2/J_1
\sim 1$. The reduced entropy and its first derivative are
illustrated as functions of $J_2/J_1$ in Fig.3 (b).  By analyzing
the first derivative $d(S_N/N)/d(J_2/J_1)$ as a function of
$J_2/J_1$, we locate two local extreme points at $J_2/J_1 \sim
0.37$ and $0.62$, respectively.  According to the above criterion,
we identify  two critical
points in the system. Besides, there is also a local maximum of
$S_N$ at $J_2/J_1$=$0.5$, which is predicted to be another
first-order QPT point according to our earlier study~\cite{Chen04}
and many new results including those to be presented later.
Interestingly, there is indeed a general consensus in previous
studies that three distinct QPT points are present around $J_2/J_1
\sim 0.38$, $0.5$ and
$0.6$~\cite{Kotov99,Sushkov01,Batista04,footnote3}. In view of the
small $4 \times 4$ cluster used in our calculation, the good
agreement of our results with the above consensus  is very
impressive.

Next we consider one of the most frustrated 2D antiferromagnets,
the checkerboard antiferromagnet, which was investigated by some
analytical and numerical approaches before~\cite{checker1}. Its
ground state has Neel long-range order at $J_{\times}/J \ll 1$,
whereas it corresponds decoupled Heisenberg chains when
$J_{\times}/J \gg 1$. At $J_{\times}/J = 1$, it is a valence bond
crystal in singlet plaquettes.  These three phases are separated
by two quantum critical points at $(J_{\times}/J)_c \sim 0.80$ and
1.10, respectively. A recent field theory study reported a new
candidate phase diagram with additional intermediate ordered
phase~\cite{Starykh05}. In Fig. 4(a), the calculated reduced
entropy and its derivative are plotted as functions of
$J_{\times}/J$. As we can see, there exist a local maximum in
$S_N$
and two local extremes in its first derivative  located
respectively at $J_{\times}/J \sim 0.93$, $J_{\times}/J \sim
0.82$, and $1.02$.
Based on our scenario, we predict that there exist eventually four
different phases divided by three transition points. The same
technique is also applied to study the hard-core bosons on the
checkerboard lattice, revealing its possible superfluid to
supersolid transition~\cite{footnote4}.

\begin{figure}[t]
\includegraphics[width=6.6cm]{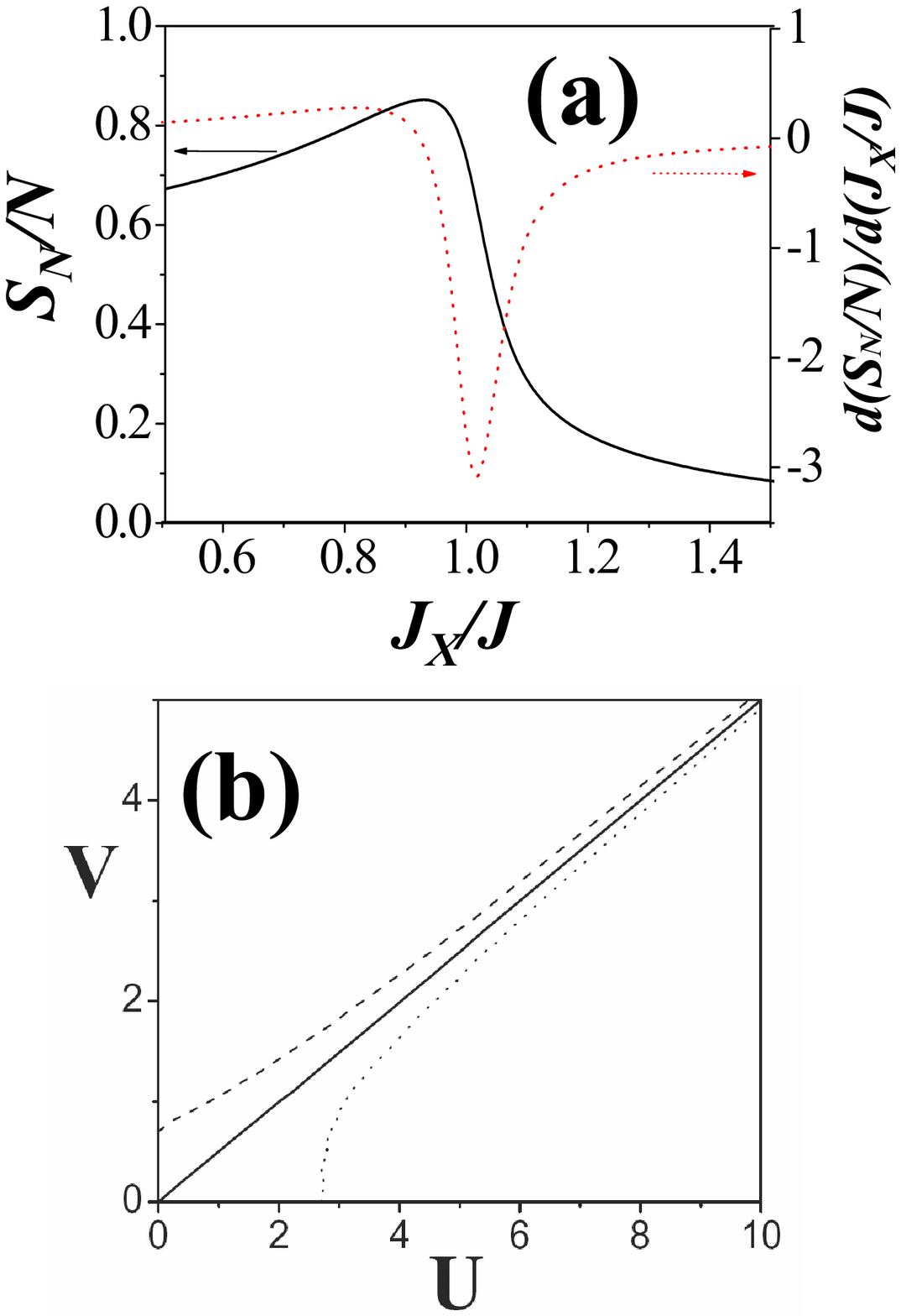}
\caption{\label{Figure 4} (a) The ground state renormalized
 reduced entropy $S_N/N$ (solid line) and its first
derivative (dotted line) versus the external parameter
$J_{\times}/J$ for the 2D checkerboard antiferromagnet model. (b)
The phase boundaries for the 1D extended repulsive Hubbard model.}
\end{figure}

A typical electron example is the 1D extended Hubbard model at half
filling (Fig.1(d)).  It is well known that there are two classical
phases: charge-density-wave (CDW) ($U/V\ll1$) and spin-density-wave
(SDW) ($U/V\gg1$) for positive $U$. The classical first-order phase
boundary lies at $U/V$=$2$~\cite{Gu04}. By employing our approach,
we can identify accurately this classical first-order boundary from
the local maxima of the reduced entropy. On the other hand, a recent
theoretical study presumably indicated the presence of a new quantum
phase bond-order-wave (BOW) at the verge of the line $U/V$=$2$, with
the QPTs between SDW/BOW and CDW/BOW appearing to be the
second-order~\cite{Sandvik04}. 
By analyzing the local extreme of the first derivative $d(S_N/N)/dV$
at fixed $U$, two critical points close to the line $U/V$=$2$ show
up. Thus we incline to predict that two new
quantum phases 
emerge around the classical phase boundary where the residual CDW
and SDW correlations may still survive in the presence of quantum
fluctuations, 
with the two new phase boundaries being the second-order
(Fig.4(b)). By increasing the lattice size (from $N$=6 to 10), the
shifts of local extreme lines of $d(S_N/N)/dV$ towards the line
$U/V$=$2$ can be clearly observed (not shown here). Thus the phase
region for BOW becomes narrower, which is consistent with a
previous investigation~\cite{Sandvik04}. It is notable that the
well-known second-order superconducting phase transition is also
revealed with the same analysis for a negative $U$.

For the above numerical finding of the relationship between the
 reduced entropy of sublattice and the transition points, particularly
from the first example, we may heuristically understand its
origin: as $N \rightarrow \infty $, the shape of the peaks or dips
around the transition points (in Figs. 3 and 4) would become
sharper~\cite{Note3}, namely  $d^2 (S_N/N)/d^2 \lambda$ (or $d
(S_N/N)/d \lambda$)  would be discontinuous at the second (first)
order transition points, resembling to the cases in thermodynamic
phase transitions.

Many QPTs  are yet to be clearly understood due to the competing
nature of highly degenerate quantum states~\cite{Sachdev}. In the
sense that the reduced entropy for a small system may capture
essentially the nonlocal feature of entanglement in a real system,
this appropriate measure of entanglement enables us to establish
its clear connection to transition points of QPTs. Therefore,
comparing with the conventional wisdoms, it is not compulsory for
us to deal with a large system to make the scaling analysis. In
our scenario, exact numerical results for a quite small correlated
system are able to disclose relevant information about QPTs.  This
is indeed a unique and superior advantage of our approach, making
it be quite promising for the future exploration of QPTs in
various physical systems, particularly if they fail to be
characterized by the conventional approach.

We thank P. Zanardi, Z.Y. Weng, Y.Q. Li, X.G. Wen, and D. Lidar for
helpful discussions. This work was supported by the RGC grants of
Hong Kong, the URC fund of HKU, and the NSFC grant(10429401).


\begin{thebibliography}{}
\bibitem{footnote1} Let us consider the three dimensional Anderson localization
model, with $H= -\sum_{\langle ij \rangle \sigma }c_{i\sigma
}^{\dagger }c_{j\sigma} + \sum_{i \sigma} \epsilon_{i} c_{i\sigma
}^{\dagger }c_{i\sigma}$. The disorder is incorporated into the
diagonal parameter $\epsilon_{i}$, which changes randomly with
 a uniform distribution $[-W/2, W/2]$ where $W$
characterizes the strength of the disorder. We here examine the
reduced entanglement and found clearly its monotonic decrease
 with respect to $W$, while its
first derivative  has a local minimum at $W_c \sim 16$ ($W_c$ is the
critical disorder strength) which corresponds to a second-order
metal-insulator transition, in excellent agreement the well-known
result.
\bibitem{footnote2} 
This feature implies that the reduced entropy of sublattice
introduced by us is an extensive quantity just like the thermal
entropy, which is distinctly from other entanglement measures
proposed before and is a reason why it can play a similar role in
QPTs as that played by the thermal entropy in thermodynamic phase
transitions.
\bibitem{footnote3} Previous studies indicated surely that the transition
from the Neel state to the spin-liquid state around $J_2/J_1 \sim
0.38$ is of  second-order, while the results for the orders around
$J_2/J_1 \sim 0.50$ and $J_2/J_1 \sim 0.62$ were likely more
obscure/debatable. Although a series expansion
study~\cite{Sushkov01} claimed a second-order transition at
$J_2/J_1 \sim 0.50$, it was indicated later that it should be the
first order~\cite{Batista04}, in agreement with our result. Also
note that Ref.\cite{Sushkov01} argued that the point $J_2/J_1 \sim
0.6$ is probably of first-order, but is very close to
second-order.
\bibitem{footnote4} Our hard-core bosons Hamiltonian reads $H =
- t \sum_{\langle ij \rangle}\left(b^{\dag}_i b^{\vphantom\dag}_j
+ b_j^{\dag} b^{\vphantom\dag}_i \right) + \sum_{\langle ij
\rangle} V (n_i-\frac{1}{2}) (n_j-\frac{1}{2})$ with the hopping
intergral $t$, nearest-neighboring repulsion $V$ and the boson
creation operator $b^{\dag}_i$. This model is eqivalent to a
spin-1/2 XXZ model, $H = \sum_{\langle ij \rangle} \left[- J_0
\left( S^x_i S^x_j+S^y_i S^y_j \right) + J_z S^z_i S^z_j\right]$
with $J_0=2t$ and $J_z=V$. We use the same technique to analysize
the reduced entropy and found a local maxima at $J_z/J_0 \sim 10$
which may correspond to a possible first-order superfluid to
supersolid transition. It is clear that in the large $J_z (V)$
limit, a novel supersolid phase may show up due to the remarkable
geometrical frustration.
\end{thebibliography}
\end{document}